\documentclass{rstransa} 

\usepackage{amsfonts}
\usepackage{amsmath}
\usepackage{amssymb}
\usepackage{booktabs}
\usepackage{caption}
\usepackage{color}
\usepackage{comment}
\usepackage{float}
\usepackage{graphicx}
\usepackage{hyperref}
\usepackage[utf8]{inputenc} 
\usepackage{subfig}
\usepackage{xspace}
\usepackage{anyfontsize} 
\usepackage{multirow} 
\usepackage{paralist}       

\newcommand{\pygbe}{\texttt{PyGBe}\xspace}

\graphicspath{{figs/}} 

\begin{document}
\title{Reproducible Validation and Replication Studies in Nanoscale Physics}

\author{
N. C. Clementi$^{1}$, L. A. Barba$^{1}$}

\address{$^{1}$Department of Mechanical and Aerospace Engineering, 
The George Washington University, Washington D.C., USA }

\subject{computer modeling and simulation, computational physics}


\corres{L.A. Barba\\
\email{labarba@gwu.edu}}

\begin{abstract}
    Credibility building activities in computational research include
    verification and validation, reproducibility and replication, and
    uncertainty quantification. Though orthogonal to each other, they are
    related. This paper presents validation and replication studies in 
    electromagnetic excitations on nanoscale structures, where the quantity
    of interest is the wavelength at which resonance peaks occur. 
    The study uses the open-source software PyGBe: a boundary element
    solver with trecode acceleration and GPU capability.
    We replicate a result by Rockstuhl et al. (2005, doi:10/dsxw9d) with a
    two-dimensional boundary element method on silicon carbide particles,
    despite differences in our method. 
    The second replication case from Ellis et al. (2016, doi:10/f83zcb) looks
    at aspect ratio effects on high-order modes of localized surface phonon-polariton
    nanostructures. The results partially replicate: the wavenumber
    position of some mode match, but for other modes they differ.
    With virtually no information about the original simulations, 
    explaining the discrepancies is not possible. 
    A comparison with experiments that measured polarized reflectance 
    of silicon carbide nano pillars provides a validation case.
    The wavenumber of the dominant mode and two more do match, but 
    differences remain in other minor modes. 
    Results in this paper were produced with strict reproducibility
    practices, and we share reproducibility packages for all, including
    input files, execution scripts, secondary data, post-processing code and
    plotting scripts, and the figures (deposited in Zenodo).
    In view of the many challenges faced, we propose that reproducible
    practices make replication and validation more feasible.

    \end{abstract}
    
    
\begin{fmtext}

\end{fmtext}

\maketitle

\section{Introduction}

Some fields of research, particularly solid mechanics and computational fluid dynamics, have a long tradition of community consensus building and established practices for verification and validation of computational models. 
Such practices are uncommon in other fields of science, especially if they have more recently become computationally intensive.
Verification and validation also become increasingly difficult when the computational models arise from many levels of mathematical and physical modeling, representing a complex system. 
In recent years, science as a whole has come to be concerned with reproducibility and replication as a new front in the continual campaign to build confidence on published findings. 
Together with formal processes of uncertainty quantification, we have now three complementary ``axes'' for building trust in science. 

The lengths to which research communities should go to conduct activities in verification and validation, reproducibility and replication, and uncertainty quantification, are highly debated. 
Some journals require articles reporting on computational results to include proof of these activities, while most do not consider these aspects at all in their review criteria. 
In this paper, we tackle a sub-field of computational physics where tradition for these confidence-building activities is scant. 
The physical setting, excitation of resonance modes in nanostructures under an electromagnetic field, relies on multiple levels of modeling, while the experimental methods are complicated by the small length scales. 
We previously developed a computational model and software (called \pygbe) that has undergone code and solution verification, but a validation opportunity had remained elusive. 
Here, we present replication studies and a validation case based on published simulation and experimental results. 
Moreover, the studies in this paper were conducted under rigorous reproducibility practices, and all digital artifacts needed to reproduce every figure are shared in reproducibility packages available in a GitHub repository and archival services. 

\section{Background and methods}\label{sec:background}

\subsection{Verification, validation, reproducibility and replication}

Verification and validation of computational models---often abbreviated V\&V and viewed in concert---have developed into a mature subject with more than two decades of organized efforts to standardize it, dedicated conferences, and a  journal. 
The American Society of Mechanical Engineers (ASME), a standards-developing organization, formed its first Verification and Validation committee (known as V\&V 10) in 2001, with the charter: 
``to develop standards for assessing the correctness and credibility of modeling and simulation in computational solid mechanics.''
It approved its first document in 2006: The Guide for Verification and Validation in Computational Solid Mechanics (known as V\&V 10-2006). 
The fact that this guide was five years in the making illustrates just how complex the subject matter, and building consensus about it, can be. 
Since that first effort, six additional standards sub-committees have tackled V\&V in a variety of contexts. 
V\&V 70 is the latest, focused on machine-learning models.
The key principles laid out in the first V\&V standard persevere through the many subsequent efforts that have operated to this day. 
They are:

\begin{compactitem}

\item[$\triangleright$] Verification must precede validation.
\item[$\triangleright$] The need for validation experiments and the associated accuracy requirements for computational model predictions are based on the intended use of the model.
\item[$\triangleright$] Validation of a complex system should be pursued in a hierarchical fashion from the component level to the system level.
\item[$\triangleright$] Validation is specific to a particular computational model for a particular intended use.
\item[$\triangleright$] Validation must assess the predictive capability of the model in the physical realm of interest, and it
must address uncertainties that arise from both simulation results and experimental data.

\end{compactitem}

\noindent
The process of \emph{verification} establishes that a computational model correctly describes the intended mathematical equations and their solutions.
It encompasses both code correctness, and solution accuracy.
\emph{Validation}, on the other hand, seeks to determine to which measure a computational model represents the physical world. 
We like to say that ``verification is solving the equations right, and validation is solving the right equations'' \cite{roache1998}. 
But in reality the exercise can be much more complicated than this sounds. 
Computational models in most cases are built in a hierarchy of simplifications and approximations, and comparing with the physical world means conducting experiments, which themselves carry uncertainties. 

As we will discuss in this paper, verification and validation in contexts that involve complex physics at less tractable scales (either very small, or very large), or where experimental methods are nascent, proceeds in a tangle of researcher judgements and path finding. 
In practice, validation activities reported in the scholarly literature often concentrate on using a stylized benchmark, and comparing experimental measurements with the results from computational models on that benchmark. 
Seldom do these activities address the key principles of pursuing validation in a hierarchical fashion from the component to the system level, and of assessing the predictive capability of the computational model accounting for various sources of uncertainties. 
Comprehensive validation studies are difficult, expensive, and time consuming. 
Often, they are severely limited by practical constraints, and the conclusions equivocal. 
Yet the computational models still provide useful insights into the research or engineering question at hand, and we build trust on them little by little.

Verification and validation align on one axis of the multi-dimensional question of when are claims to knowledge arising from modeling and simulation justified, credible, true \cite{winsberg-2010}.
Two other axes of this question are: reproducibility and replication, and uncertainty quantification (UQ).
Uncertainty quantification uses statistical methods to give objective confidence levels for the results of simulations. 
Uncertainties typically stem from input data, modeling errors, genuine physical uncertainties, random processes, and so on. 
A scientific study may be reproducible, the simulations within it undergone V\&V, yet the results are still uncertain. 
Building confidence in scientific findings obtained through computational modeling and simulation entails efforts in the three ``axes of truth'' described here.

Reproducibility and replication (we could call it R\&R) preoccupy scientific communities more recently. 
Agreement on the terminology, to begin with, has been elusive \cite{barba2018}. 
The National Academies of Science, Engineering and Medicine (NASEM) released in May 2019 a consensus study report on Replicability and Reproducibility in Science \cite{NASEM2019} with definitions as follows.
``Reproducibility is obtaining consistent results using the same input data, computational steps, methods, and code, and conditions of analysis.
Replicability is obtaining consistent measurements or results, or drawing consistent conclusions using new data, methods, or conditions, in a study aimed at the same scientific question.''
According to these definitions, reproducibility demands full transparency of the computational workflow, which at the very least means open code and open data, where `open` means shared at time of publication (or earlier) under a standard public license. 
This condition is infrequently satisfied.
The NASEM report describes how a number of systematic efforts to reproduce computational results have failed in more than half of the attempts made, mainly due to inadequately specified or unavailable data, code and computational workflow \cite{moraila-etal-2014,iqbal-etal-2016,chang-li2018,stodden-etal-2018}. 
Recommendation 4-1 of the NASEM report states that 
``to help ensure reproducibility of computational results, researchers should convey clear, specific, and complete information about any computational methods and data products that support their published results in order to enable other researchers to repeat the analysis, unless such information is restricted by nonpublic data policies. That information should include the data, study methods, and computational environment'' \cite{NASEM2019}.

Although it may seem evident that running an analysis with identical inputs would result in identical outputs, this is sometimes not true. 
For example, the combination of floating-point representation of numbers and parallel processing means that running the same software with the same input data may give different numerical results. 
In some research settings, it may make sense to relax the requirement of bitwise reproducibility and settle on reproducible results within an accepted range of variation (or uncertainty). 
This can only be decided, however, after fully understanding the numerical-analysis issues affecting the outcomes---and the risk associated with an uncertainty range. 
Researchers using high-performance computing thus recognize that when different runs with the same input data produce slightly different numeric outputs, each of these results is equally credible, and the output must be understood as an approximation to the correct value within a certain accepted uncertainty.

Beyond the particulars of high-performance or parallel computing, 
a number of factors can contribute to the lack of reproducibility in research. 
In addition to lack of access to non-public data and code, the NASEM report (of which Barba is a co-author) lists the following contributors to lack of reproducibility:

\begin{compactitem}

\item[$\triangleright$] Inadequate record-keeping: the researchers did not properly document all relevant digital artifacts and steps followed to obtain the results, the details of the computational environment, software dependencies, and/or identifiers and metadata for data products.
\item[$\triangleright$]  Lack of transparency: the researchers did not transparently report, using standard public licenses, an archive with all relevant digital artifacts necessary to reproduce the results.
\item[$\triangleright$]  Obsolescence of the digital artifacts: over time, the digital artifacts in the research compendium are compromised because of technological breakdown and evolution or lack of continued curation.
\item[$\triangleright$]  Flawed attempts to reproduce others' research: the researchers who attempted to reproduce the work lacked expertise or failed to correctly follow the research protocols.
\item[$\triangleright$]  Barriers in the culture of research: lack of resources and incentives to adopt reproducible and transparent research across fields and researchers.

\end{compactitem}

\noindent
Improving computational reproducibility hinges on capturing and sharing information about the computational environment and steps required to collect, process, and analyze data.
Scientific workflows represent a complex flow of data products through various steps of collection, transformation, and analysis to produce an interpretable result. 
Capturing provenance of the result is increasingly difficult to do using manual processes, and automation is key. 
With regards to software management, version-control systems are used to automatically capture the history of all changes made to the source code of a computer program. 
This creates a history of changes and allows the developers to better understand the code and to identify possible problems or errors.
Recent technological advances in version control, virtualization, computational notebooks, and automatic provenance tracking have the potential to simplify reproducibility, and tools have been developed that leverage these technologies.
Still, many questions remain unanswered both to understand the gaps left by existing tools and to develop principled approaches that fill those gaps. 

Replication of scientific findings is key for building trust in them. 
It is often difficult to attain, for many reasons, not least because deciding when two scientific findings are \emph{consistent} is tangled in researcher judgements and inevitable constraints. 
The NASEM report lists the following factors affecting the replicability of findings: 

\begin{compactitem}

\item[$\triangleright$] the complexity of the system under study;
\item[$\triangleright$] understanding of the number and relations among variables within the system under study;
\item[$\triangleright$] the ability to control the variables; 
\item[$\triangleright$] levels of noise within the system (or signal to noise ratios);
\item[$\triangleright$] the mismatch of scale of the phenomena and the scale at which it can be measured; 
\item[$\triangleright$] stability across time and space of the underlying principles; 
\item[$\triangleright$] fidelity of the available measures to the underlying system under study (e.g., direct  or indirect measurements); and
\item[$\triangleright$] prior probability (pre-experimental plausibility) of the scientific hypothesis.

\end{compactitem}

\noindent
Fields of study that have been at the center of the `replication crisis` commotion tend to be characterized by their complexity, intrinsic variability, or inability to control variables, e.g., psychology. 
But in many areas of modern technology we face similar challenges to control variables or disentangle many interacting effects. 
In this paper, we tackle replication and validation of computational models in nanoscale physics, where certainly the systems are complex, variables difficult to control, and signals subject to noise.

\subsection{Description of the PyGBe software}

Our research group has been developing PyGBe---pronounced \emph{pigb\={e}}---for several years. 
It was first written for biomolecular-electrostatics calculations using a continuum model of proteins in water or an ionic solvent. 
The computational model applies a boundary integral form of the governing Poisson-Boltzmann and Poisson equations, to obtain the electrostatic potential and its normal derivative on the molecular surface. 
In the implicit-solvent model, this information is used to compute the quantity of interest: solvation energy, which is diagnostic for questions of protein binding affinity, protein-surface interactions, and others. 
Biomolecules are modeled as dielectric cavities inside an infinite continuum solvent, and the partial differential equations are solved via a boundary element method, leading to a dense linear system of equations. 
PyGBe uses a fast-summation algorithm via a Barnes-Hut treecode \cite{BarnesHut1986}, and accelerates the computationally intensive parts of the code on Nvidia GPU hardware using CUDA kernels in the treecode, interfacing with PyCUDA \cite{klockner2012pycuda}. 
Other portions of the code are written in C$++$, wrapped using \texttt{swig}, to extract more performance \cite{CooperETal2016}. 
These features allow PyGBe to handle problems with half a million boundary elements or more. 

In more recent work, we expanded the capabilities of PyGBE to applications of computational nanoplasmonics for biosensing \cite{ClementiETal2019}. 
Applying the long-wavelength limit, one can model the phenomenon of localized surface plasmon resonance (LSPR) via electrostatics, instead of the full Maxwell equations. 
This phenomenon is harnessed in nanosensors for detecting with high sensitivity the presence of biomolecules through shifts in resonance frequency. 
In LSPR, an electromagnetic wave excites the free electrons on the surface of a metallic nanoparticle. 
The name given to these electron-cloud vibrations is plasmons. 
In this setting, they resonate with the incoming field, and the energy is either absorbed by the nanoparticle or scattered in all directions, causing what's called extinction.
Since the resonance frequency depends on the dielectric environment, the change produced by a biomolecule approaching the metallic nanoparticle results in a frequency shift, and a means of detecting its presence.
The electrostatic approximation allows using the methods implemented in PyGBe, after substantial code development to permit using complex-valued permittivities, and to incorporate an external electric field into the model. 
These changes included re-writing the Krylov iterative solver to work with complex numbers, and splitting the treecode evaluation into real and imaginary parts. 
New functions were added to read from configuration files the data describing the electric field, to compute the dipole moment, and to compute the final quantity of interest: extinction cross-section. 
We reported a major new release of the software in 2017 \cite{ClementiETal2017}, and later  presented the mathematical formulation for electromagnetic scattering in the long-wavelength setting and its associated continuum and discretized integral equations, and results including verification activities and sensitivity calculations on a biosensor model \cite{ClementiETal2019}. 
For verification, we conducted grid-convergence studies in two settings.
In the first, we set up a computational model for a spherical silver nanoparticle in a constant electric field, leading to an analytical solution for the extinction cross-section. 
The second case does not have a closed form solution: a spherical nanoparticle with a nearby protein (analyte), under an electric field. 
To estimate relative errors in the extinction cross-section, in this case, we made use of the method of Richardson extrapolation.
These verification activities build our confidence on the computational model---moreover, our work was published following careful reproducibility practices, including the release of reproducibility packages for all results. 
Nevertheless, the gold standard of confidence in the predictive capability of the computational model comes from validation: our quest for an opportunity to conduct validation studies with PyGBe in these settings led us through the twisted path that we report in this paper.

\subsection{Physics context for this work}

In recent years, polar dielectric crystals such as Silicon Carbide (SiC) became recognized as an alternative to plasmonic metals in many technologies, including biosensors. 
They manifest oscillations of lattice-bound charges, called surface phonon polaritons, in the mid- to long-wave infrared range with low optical losses.
Nanostructures made of these materials offer sensing capabilities, described by their \emph{figure of merits}, that are unattainable with plasmonic metals. 
The figure of merits of a nanoparticle is defined as the ratio between the sensitivity and the width of the resonance peak at mid-height \cite{otte-etal-2012}. 
In turn, sensitivity is the shift in the resonance position divided by the change in the refractive index: 
$S = \Delta \lambda / \Delta n$.
The dielectric function of polar dielectrics has a negative real part and a small imaginary part, in the mid to long infrared regime. 
This dielectric behavior is similar to that observed in metals like silver in the blue part of the wavelength spectrum. 
In plasmonics, when illuminating a small particle made of metallic materials, we will observe that certain wavelengths excite a surface plasmon. 
The main difference with polar dielectrics like SiC is that for this material the frequency of the incoming light matches instead the resonance frequency of the Si and C sub-lattices \cite{caldwell2015,rockstuhl2005}. 
This excitation leads to a strong extinction cross-section at the resonance wavelength, and an enhanced near-field amplitude. 
These behaviors can be modeled with the same approaches used for localized surface plasmon resonance, and when the wavelength is much larger than the size
of the nanoparticle, we can again apply the electrostatic approach implemented in \pygbe \cite{ClementiETal2017, ClementiETal2019}.

When studying both surface phonon polaritons and surface plasmon resonances, one can analyze the spectrums by measuring different quantities. 
We can measure scattering cross-section, extinction cross-section (scattering plus absorption), as well as reflection. 
These different approaches are comparable since the \textit{quantity of interest} is the wavelength (frequency) at which the resonance modes happen. 
Whether we measure reflection or extinction cross-section, the wavelength (frequency) at which the peaks happen remains the same. 
Throughout this work, we will concentrate on studying the wavelength (frequency) at which the resonance modes occur, 
and we aim to replicate results from Rockstuhl et al.\ 2005 \cite{rockstuhl2005} and from Ellis et al.\ 2016 \cite{ellis2016}, 
and to validate our software against experimental results from Ellis et al.\ 2016 \cite{ellis2016}.

\section{Results} \label{sec:results}
All the results reported in this paper were obtained using the \pygbe software \cite{ClementiETal2017},
with the version at commit \href{https://github.com/pygbe/pygbe/tree/e1f3650b0c99cab99dcfe5372200d3a1534eddfe}{34eddfe} in the history.
The software GitHub repository contains a Dockerfile to create the container image where we ran the simulations. 
The manuscript GitHub repository is separate from the software repository, and can be found at \url{https://github.com/barbagroup/pygbe_validation_paper}.
It contains the reproducibility packages for all results, as described in sub-section \ref{sec:reprod} and elsewhere in the paper.
We used a lab workstation for all simulations, built from parts.
Hardware specifications are as follows: 
\begin{compactitem}
  \item[$\triangleright$] CPU: Intel Core i7-5930K Haswell-E 6-Core 3.5GHz LGA 2011-v3
  \item[$\triangleright$] RAM: G.SKILL Ripjaws 4 series 32GB (4 x 8GB)
  \item[$\triangleright$] GPU: Nvidia Tesla K40c (with 12 GB memory)
\end{compactitem}

\paragraph{Solver parameters:} We used a GMRES exit tolerance of $1\times10^{-6}$ in the iterative solver for all our simulations. Details on the treecode and integration parameters (and others)
can be found within input files included in the manuscript GitHub repository as part of the reproducibility packages (``repro-packs'').

\paragraph{Run times:} We report detailed time logs together with the data of our simulations, as part of the repro-packs to accompany this paper. Here is a brief report.
For the result of the replication of Figure 14 of Rockstuhl et al., sub-section \ref{sec:replication1}, the total wall-clock time for producing each curve was approximately 11.6 hours, which is the result of computing
the extinction cross section for 175 wavelength cases, giving $\approx$4 minutes per run. In the case of the validation and replication of Figure 2a of Ellis et al., sub-section \ref{sec:replication2}, 
the total wall-clock time for producing the curve was approximately 2.3 hours, which is the result of computing
the extinction cross section for 208 wavelength cases, giving $\approx$40 seconds per run.

\subsection{Replication of results from Rockstuhl, et al., 2005}\label{sec:replication1}

The work of Rockstuhl et al.\cite{rockstuhl2005} studies the phonon-polariton response of silicon carbide (SiC)
nanoparticles using a two-dimensional (2D) boundary element method, 
previously developed in their group \cite{rockstuhl2003}. 
They analyzed ``cylindrical particles'' (where they extend in the third dimension to infinity) made of 6H-SiC, with varying cross-sections.
We decided to attempt to replicate one of the results presented on Fig.14 of their paper:
the scattering cross-section of a SiC rectangular cylinder for three different aspect ratios. 
To be well within our quasistatic approach ($\lambda > d$ where $d$ is the characteristic
dimension of the geometry), we chose the case with $a=672$ nm, $b=328$ nm.

\subsubsection{Differences in method, mesh and dielectric data}

\paragraph{Method:} The main difference between the original simulations and ours is that they solve a 2D problem with the 
full Maxwell equations, while we solve a 3D problem with the electrostatic approximation. 
We lack any information about their code implementations, discretization schemes, or solver.  
We computed extinction cross-section (scattering plus absorption) while Rockstuhl et al.\ present only scattering cross-section. 
But, since the quantity of interest is the wavelength at which the resonance modes occur, the results can be compared.

\paragraph{Mesh:} Rockstuhl et al.\ did not provide any details regarding the discretization of the geometries or 
parameters involved in the simulations.
We performed a grid-independence study as a form of solution verification, and to ensure that we are 
minimizing errors due to discretization. Rockstuhl et al.\ used a 2D geometry (infinite third dimension), while we treat the geometry in its full 3D representation.

\paragraph{Dielectric data:} The study uses 6H-SiC as material, and the authors obtained their data from a a source that we
were not able to replicate. Instead we are using experimental data for 4h-SiC that was provided to us 
via private communications from the authors of Ellis et al.\cite{ellis2016}.  

\subsubsection{Grid-independence study}\label{sec:independence}

\begin{figure}
    \centering
    \includegraphics[width=0.75\textwidth]{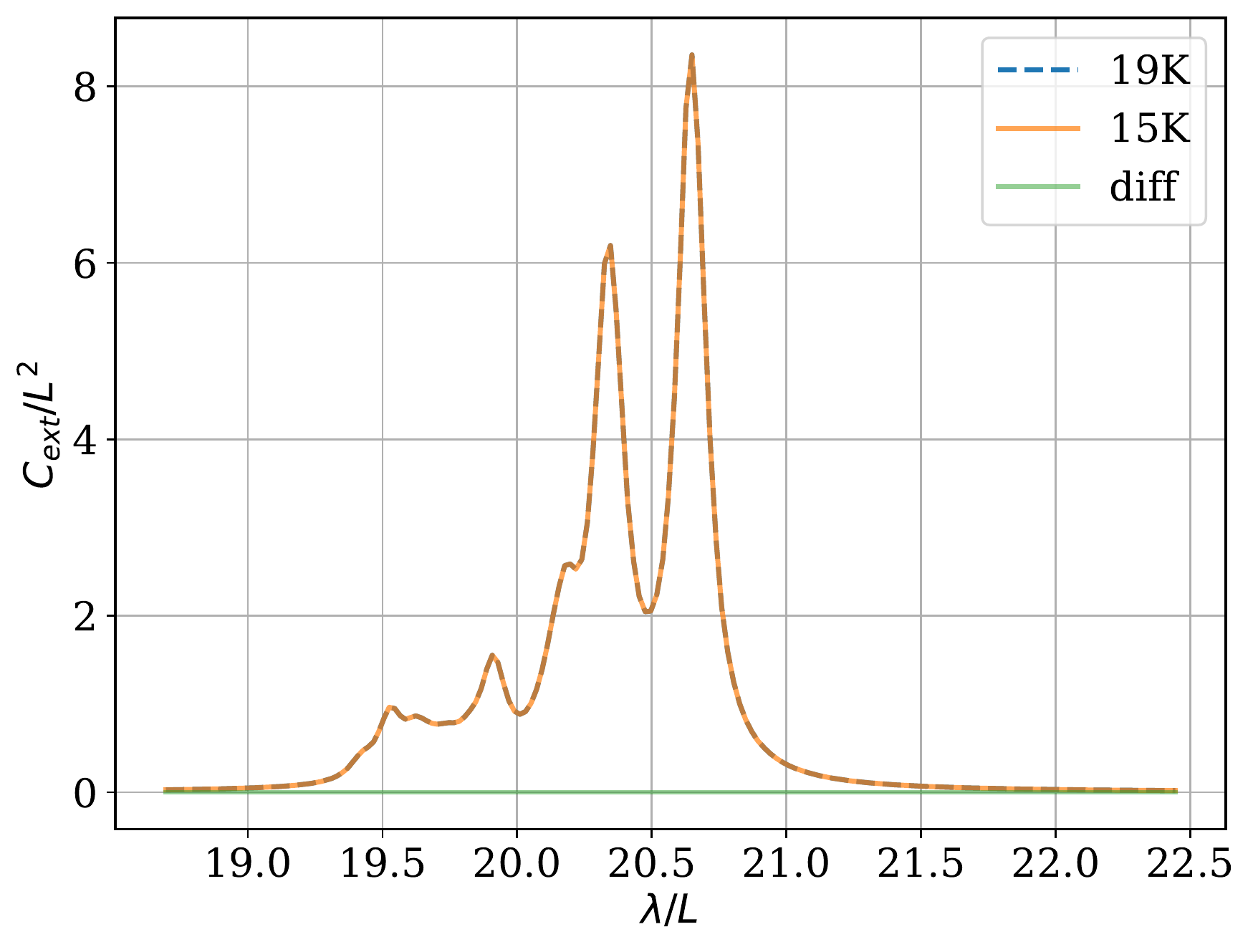} 
    \caption{Grid-independence study for a SiC cube of side $L=535$ nm submerged in air under a constant 
    electric field in the $z$-direction. The curves represent the extinction cross-section divided by $L^2$ 
    as a function of wavelength divided by $L$, for two mesh sizes as shown in the legend.}
    \label{fig:cube535}
 \end{figure}

We performed a grid-independence study on a SiC cube of side $L=535$ nm submerged in air, under a 
constant electric field aligned with the $z$-axis (a similar setup to the square cylinder on Fig. 18 of 
Rockstuhl et al.\cite{rockstuhl2005}). 
Due to the nature of the geometry and its sharp edges, it was challenging to see proper convergence. 
We have completed grid-convergence analysis for this type of physics in a previous work, but using spherical geometries \cite{ClementiETal2019}. 
With that experience, and given the difficulty caused by sharp edges, we are content with studying grid-independence here instead.
Figure \ref{fig:cube535} shows the grid-independence study, using meshes with  15,552 triangles (density $9.05\times10^{-4}$ per $\text{\AA}$ squared)
 to 19,200 triangles (density $1.11\times10^{-5}$ $\text{\AA}$ squared):
 the computed results show no discernible difference.

It is worth noting that the extinction cross-section curve in Figure \ref{fig:cube535} has extra peaks 
compared to the results of Figure 18 of Rockstuhl et al., due to the three-dimensional effects captured in our simulation and the sharp 
edges, the latter effect also being mentioned by Rockstuhl et al. The 3D effects will be approached
in the following results, where we attempt a replication of one of the results on Figure 14 of Rockstuhl et al. 

\subsubsection{Replication of Figure 14 (case a1) of Rockstuhl et al., 2005}

\begin{figure}
    \centering
    \includegraphics[width=0.45\textwidth]{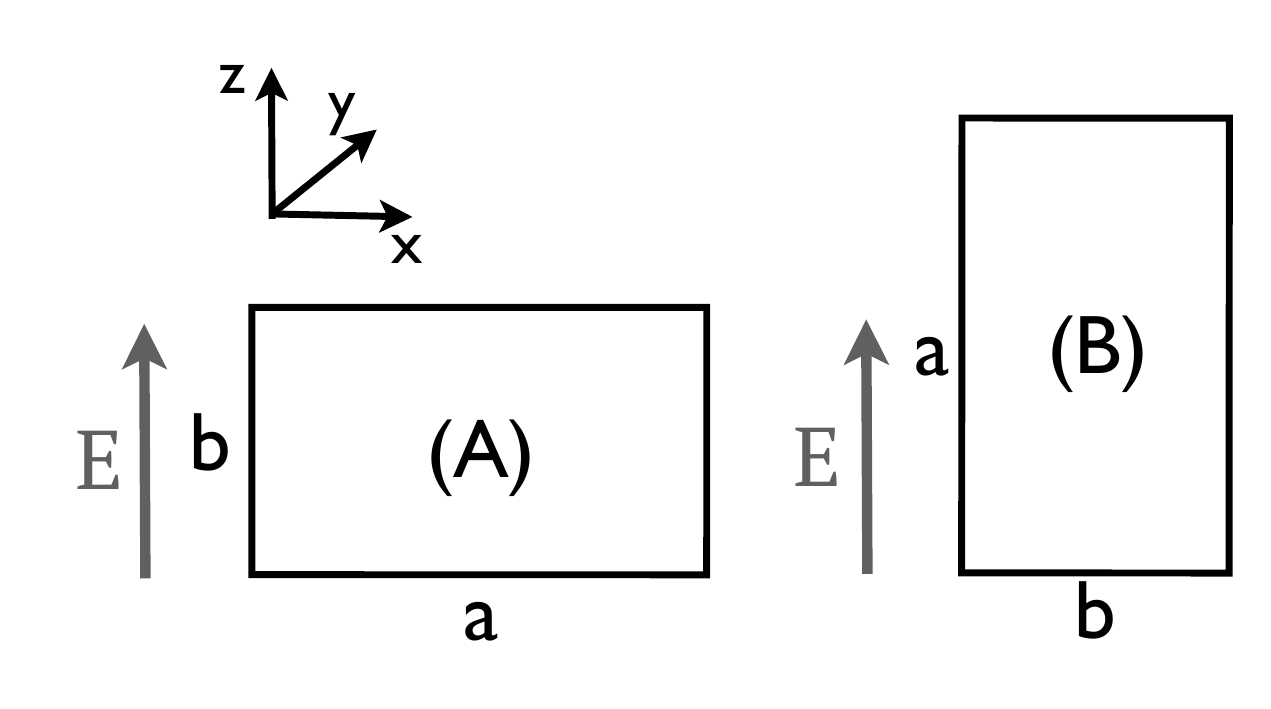} 
    \caption{Configurations for the simulations corresponding to Fig. 14 of Rockstuhl et al., 2005.}
    \label{fig:rectangle_sketch}
\end{figure}

We chose to replicate a result of Rocksuthl et al.\ presented in Figure 14: 
the case where $a=672$ nm 
and $b=328$ nm, since these dimensions are well within the limits of the quasistatic approximation 
used in \pygbe. Rockstuhl et al.\ present the normalized scattering cross-section of a SiC rectangular 
cylinder, and they perform simulations for two different setups, sketched in Figure \ref{fig:rectangle_sketch}. In 
their Figure 14 (left), they have the wave vector (illumination) along the long 
side of the geometry, which means that the electric field is parallel to the short side of the rectangle, like in 
Figure \ref{fig:rectangle_sketch} (B). Following a similar analysis, in Figure 14 (right) of Rockstuhl et al., they have the wave 
vector (illumination) along the short side of the geometry, which means that the electric field is parallel to the 
long side of the rectangle, like in Figure \ref{fig:rectangle_sketch} (A). 

\begin{figure}
    \centering
    \subfloat{\includegraphics[width=0.48\textwidth]{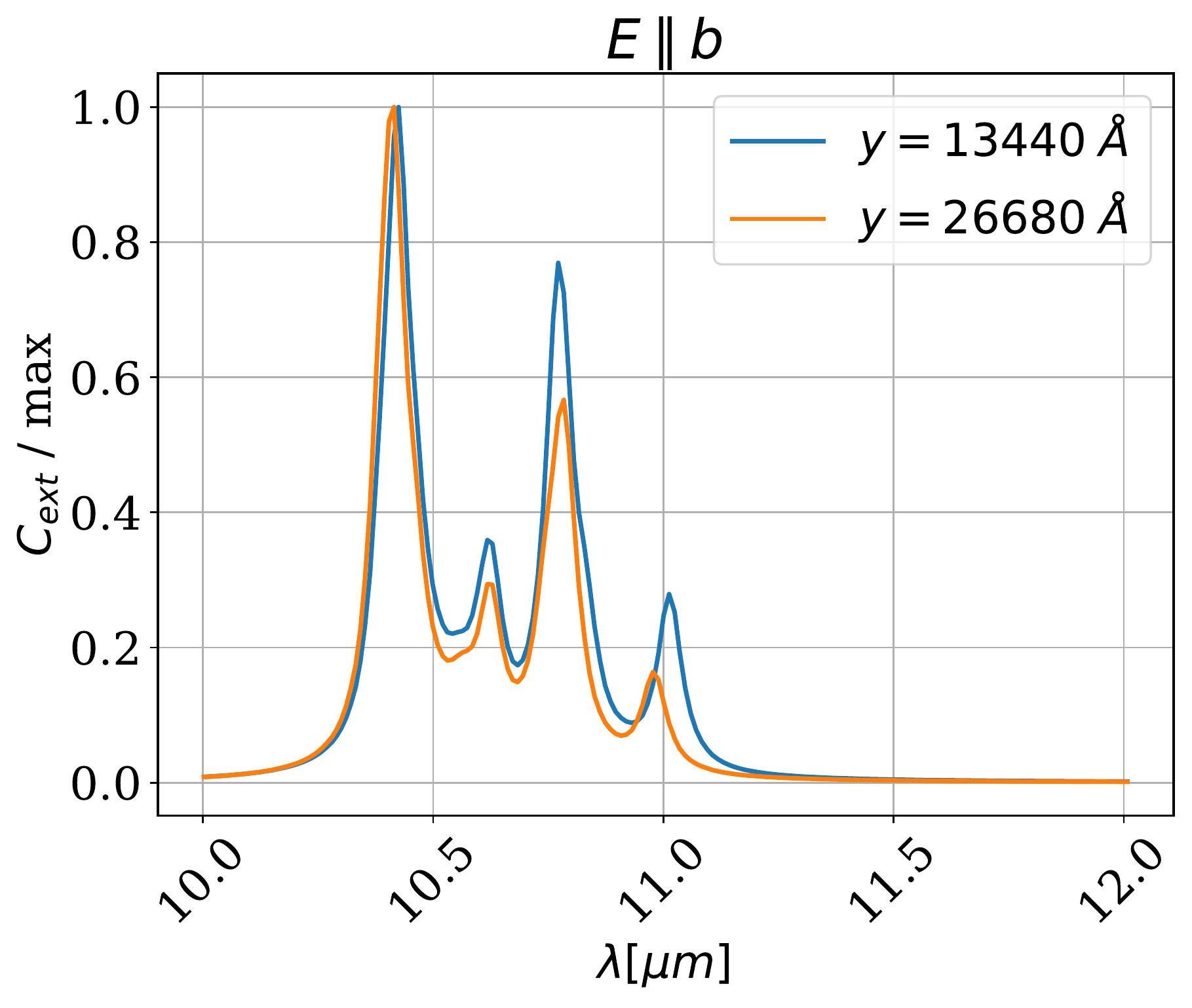}}
    \subfloat{\includegraphics[width=0.48\textwidth]{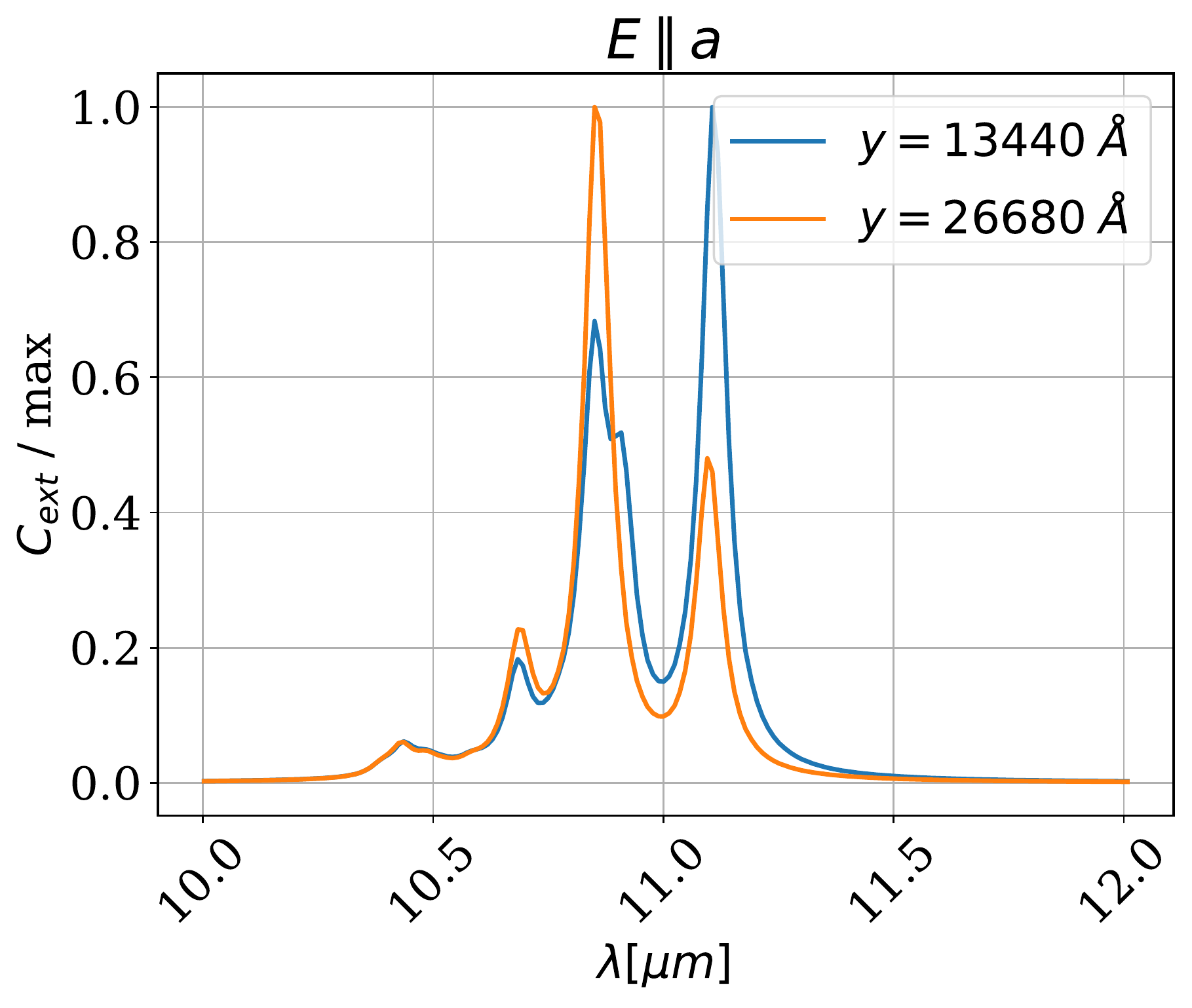}} 
    \caption{Effect of the elongation of the third dimension ($y$) on the 
        extinction cross-section of a rectangular prism of SiC of dimensions $a=672$ nm 
        and $b=328$ nm, submerged in air and under a constant electric field 
        parallel to the $z$-axis. The left plot corresponds to a configuration such that the electric 
        field is parallel to $b$ (configuration (A) on Figure \ref{fig:rectangle_sketch}), while the 
        right plot corresponds to a configuration such that the electric field is 
        parallel to $a$ (configuration (B) on Figure \ref{fig:rectangle_sketch}.}
    \label{fig:ext_y_14}   
 \end{figure}

The constraints of mesh generation mean that we have only loose control on triangle density. Our meshes for the rectangular prism all have densities like in the coarse case of the grid-independence study of sub-section \ref{sec:independence}, or finer.
For the third dimension, we needed to choose a value that represents ``infinity.'' To achieve this, we studied the effects of 
elongating the cylinder to the third dimension. In Figure \ref{fig:ext_y_14}, we present the results for two different
values of the length in the third dimension, $y=1344$ nm ($2\times a$) and $y=2688$ nm ($4\times a$). We see that as we make $y$ longer, the 
intensity of some peaks decreases, due to the fact that some of the peaks are associated to this component. Therefore, 
from now on we use $y=2688$ nm. 

\begin{figure}
    \centering
    \subfloat{\includegraphics[width=0.48\textwidth]{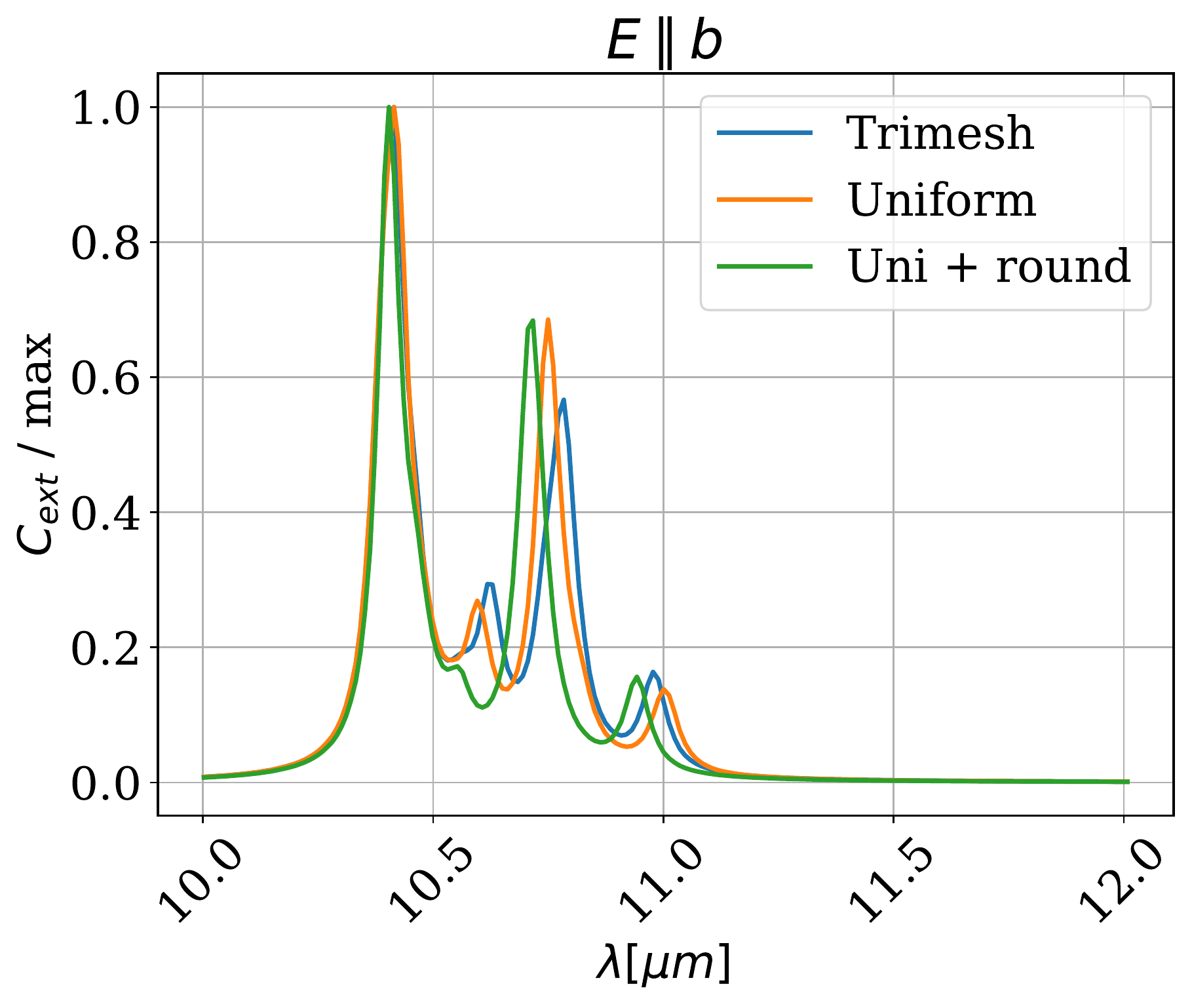}}
    \subfloat{\includegraphics[width=0.48\textwidth]{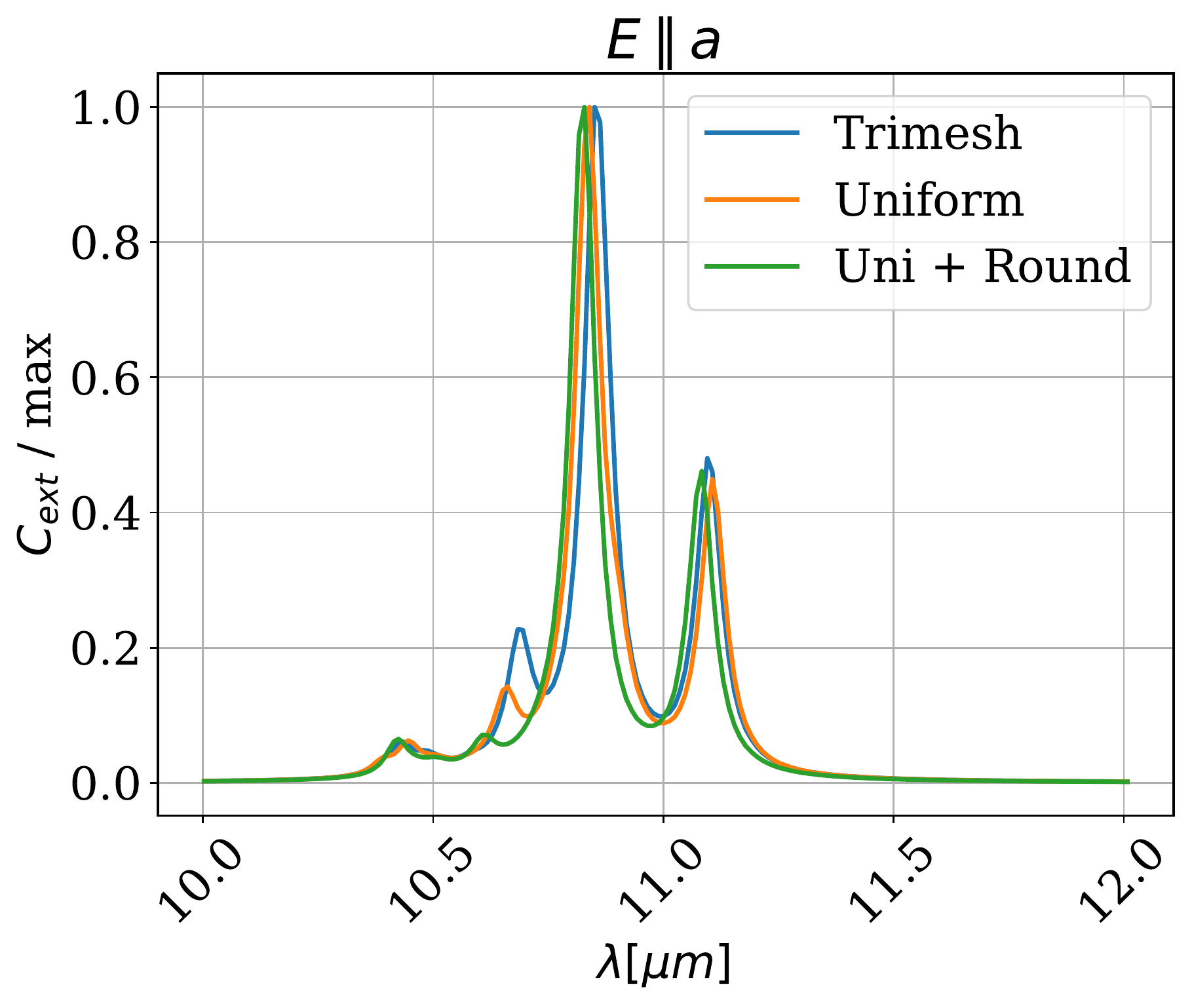}}
    \caption{Effect of uniformity of the mesh and roundness of the edges on the 
    extinction cross-section of a rectangular prism of SiC of dimensions $a=672$ nm, 
    $b=328$ nm and $y=2688$ nm, submerged in air and under a constant electric field 
    parallel to the $z$-axis.}
    \label{fig:tri_reg_round_14}
 \end{figure}

To generate the meshes for these simulations, we initially used the open source software Trimesh 
(\url{https://github.com/mikedh/trimesh}), but we realized that it was not producing a 
uniform mesh and that it was not possible to obtain regular triangles with the functions 
available when having a prism. To overcome this, we created our own mesh using Python scripts,
and obtained uniform meshes. We wanted to study the effect of a uniform mesh as well as the effect
of rounding the edges---Rockstuhl et al.\ mentioned rounded edges
as a factor that introduces extra peaks on the response. We were unable to control the 
roundness as a function of arc of curvature or the dimensions of the rectangular prism, so we 
decided to use the default settings on Trimesh. 
For reproducibility purposes, we provide all the mesh files, as described in sub-section \ref{sec:reprod}.
Figure \ref{fig:tri_reg_round_14} shows the 
results of the effect of uniformity and roundness. One can see that the second peak is not
present in the green curve, which can be attributed to the effect of the roundness. This is 
consistent with the results in Rockstuhl et al. 

\begin{figure}
    \centering
    \subfloat{\includegraphics[width=0.48\textwidth]{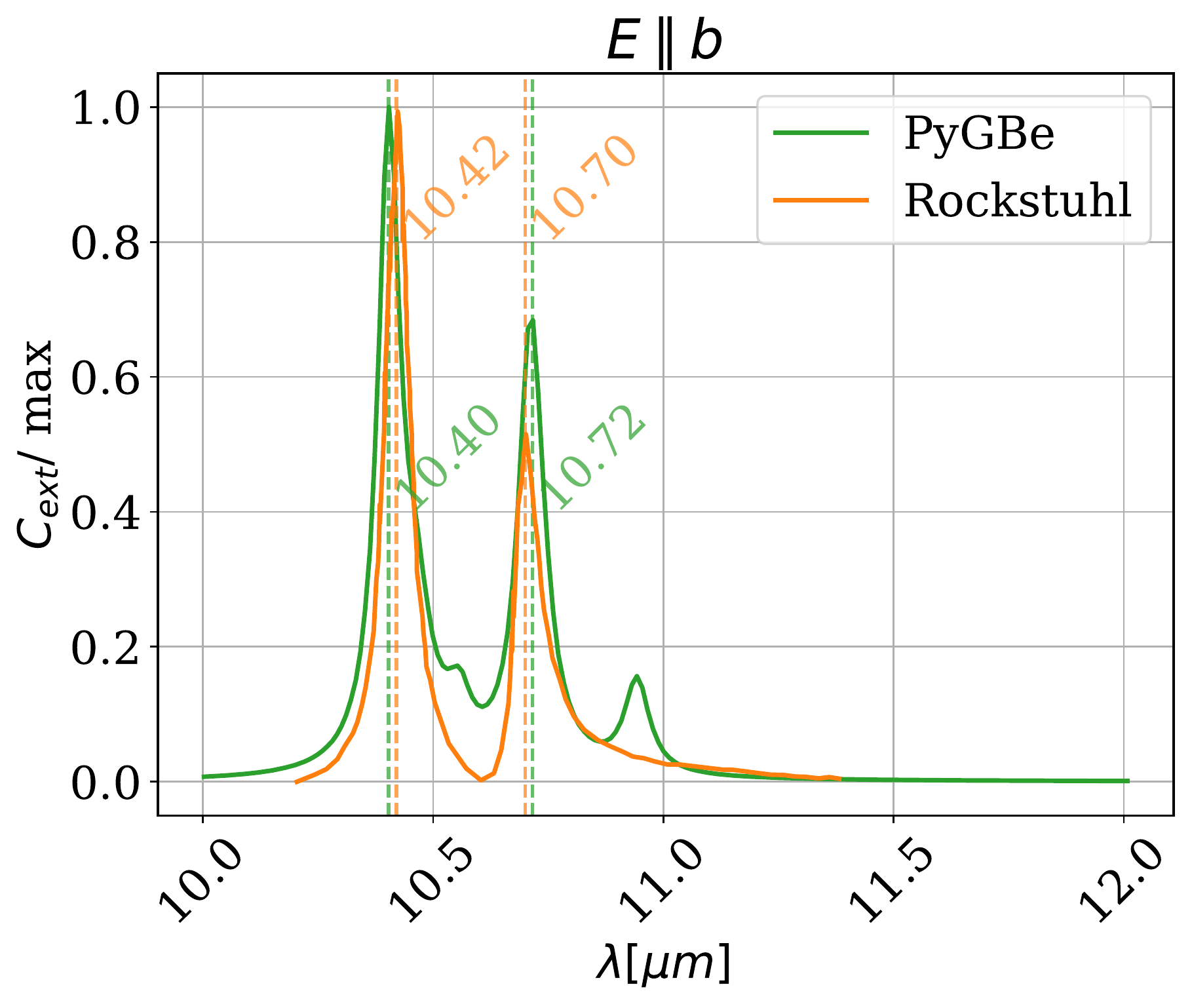}}
    \subfloat{\includegraphics[width=0.48\textwidth]{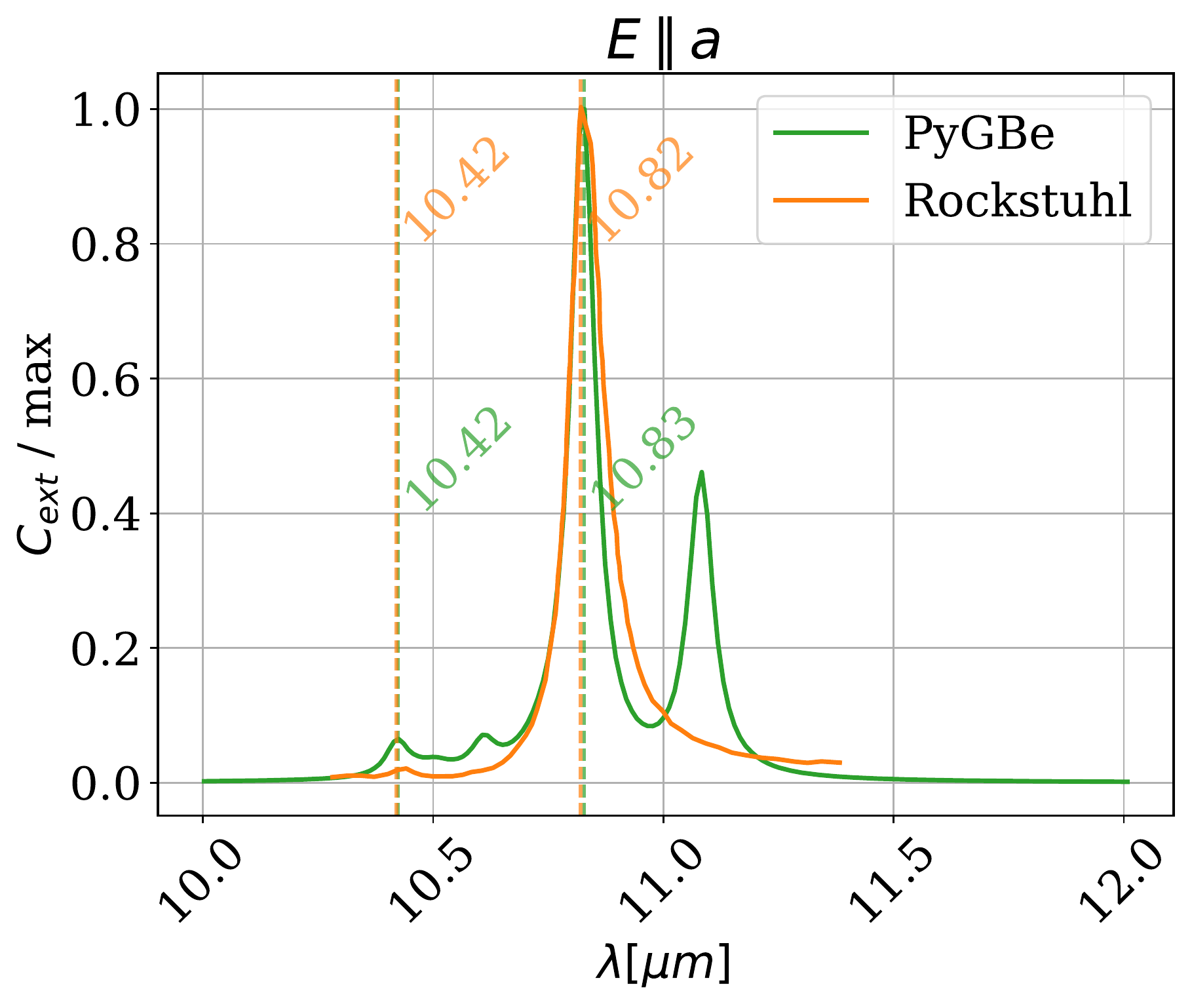}} 
    \caption{Replication of the results in Figure 14 of Rockstuhl et al., 2005. Extinction cross-section of a
    rectangular prism of SiC of dimensions $a=672$ nm, $b=328$ nm and $y=2688$ nm, submerged
    in air and under a constant electric field parallel to the $z$-axis.}
    \label{fig:rep_14}
 \end{figure}

Once we have found the ``best'' possible geometry construction, we show how our results 
(green curve in Figure \ref{fig:tri_reg_round_14}) compare 
with the original results from Rockstuhl's Figure 14. We obtained data from Rockstuhl's curves by hand using 
the WebPlotDigitizer (\url{https://apps.automeris.io/wpd/}). The replication results are 
presented in Figure \ref{fig:rep_14}: the main resonance peaks presented in Rockstuhl et al.\ are 
closely matched. While we still have the presence of a third peak in our results, we conjecture 
that this is a consequence of the 3D effects in our geometry.

 \subsection{Replication of results from Ellis et al., 2016, and validation}\label{sec:replication2}

The work of Ellis and coworkers \cite{ellis2016} studies the aspect-ratio evolution of high-order
modes in localized surface phonon-polariton nanostructures. They study the
excitation of multipolar localized surface phonon polaritons (SPhP) resonances, by measuring
and computing polarized reflectance on 4H-SiC pillars of fixed height ($H=950$ nm), fixed 
width ($W=400$ nm) and varied length ($L=400$--$4800$ nm). These pillars are patterned on a square 
grid with a pitch $P=L+500$ nm to reduce coupling. In both their simulations and experiments, they 
measured polarized reflectance with the incident polarization 
oriented parallel or perpendicular to the long axis of the pillars.  

Our first aim was to replicate the computational result presented in Figure S4 of their supplementary 
material, corresponding to the black curve on their plot. In this 
figure, they show simulation results for the resonance spectral position of the lower-frequency 
mode when having parallel polarization ($E^{\parallel}_{100}$), with an incidence angle of 22$^\circ$.
We attempted to replicate this result since the gap between the pillars 
is 5000 nm (10 times larger than in the other cases), diminishing the effects of coupling, which makes it 
a better candidate to replicate using \pygbe. 
In our calculations, the setup consists of a single pillar, with no substrate.
Our second attempted replication is  
for the results presented in Figure 2a, corresponding to reflectance measurements across wave number
for pillars with aspect ratio $AR=4$, angle of incidence of 22 degrees, and incoming polarization parallel to the 
length of the pillars. For this case, the authors also reported experimental results that we will use for validation of our solver. 

\subsubsection{Differences in method, mesh and dielectric data}

\paragraph{Method:}
Ellis et al.\ ran experiments using reflectance spectroscopy and they computed the solution of
Maxwell's equations via the RF package of the finite element solver in the commercial software COMSOL. 
In their simulations, 
they used one pillar over a portion of substrate, with periodic boundary conditions to represent an array of 
pillars and their interactions. 
We use the boundary element method in the quasistatic approximation, which is suitable in this case 
since the pillar's size is small compared to the wavelengths involved in the simulations: 
which are in the range 10000--12500 nm. 
We measure extinction cross section, which will express as peaks instead of dips (shown in the reflection plots of Ellis et al.). 
The intensity of the peaks is not comparable, but we are looking to match the wave number (quantity of interest) at which these events happen. 

\paragraph{Mesh:}
For the case with aspect ratio $AR=4$, we have a non-uniform triangular surface mesh ($N=4398$) provided to us by the authors of 
Ellis et al., which we used for validation and replication of Figure 2a of their paper. However, for the replication
of Figure S4 of the supplementary material, we needed the remaining aspect ratio meshes. We generated uniform meshes using our
Python script, and we determined the density of these meshes by comparing the extinction cross section for the case with $AR=4$ of 
our mesh and the one provided by Ellis et al. We noticed that using around double the elements ($N=8564$) than in the original mesh and 
rounding the edges, the relative errors for the extinction cross section were, on average, smaller than 3$\%$, and the variation on the wavelength peak
position was smaller than 1 cm$^{-1}$. This analysis led us to a mesh density of $\approx \; 1.7 \times10^{-5}$ triangles per Angstrom-squared to create the meshes for 
the aspect ratio variation study.

\paragraph{Dielectric data:}
The dielectric data for the simulations was given to us by the authors of the paper via a private communication, 
and corresponds to experimental values of the dielectric. 

\subsubsection{Replication of Figure S4 in the supplementary materials of Ellis et al., 2016}

\begin{figure}
    \centering
    \includegraphics[width=0.45\textwidth]{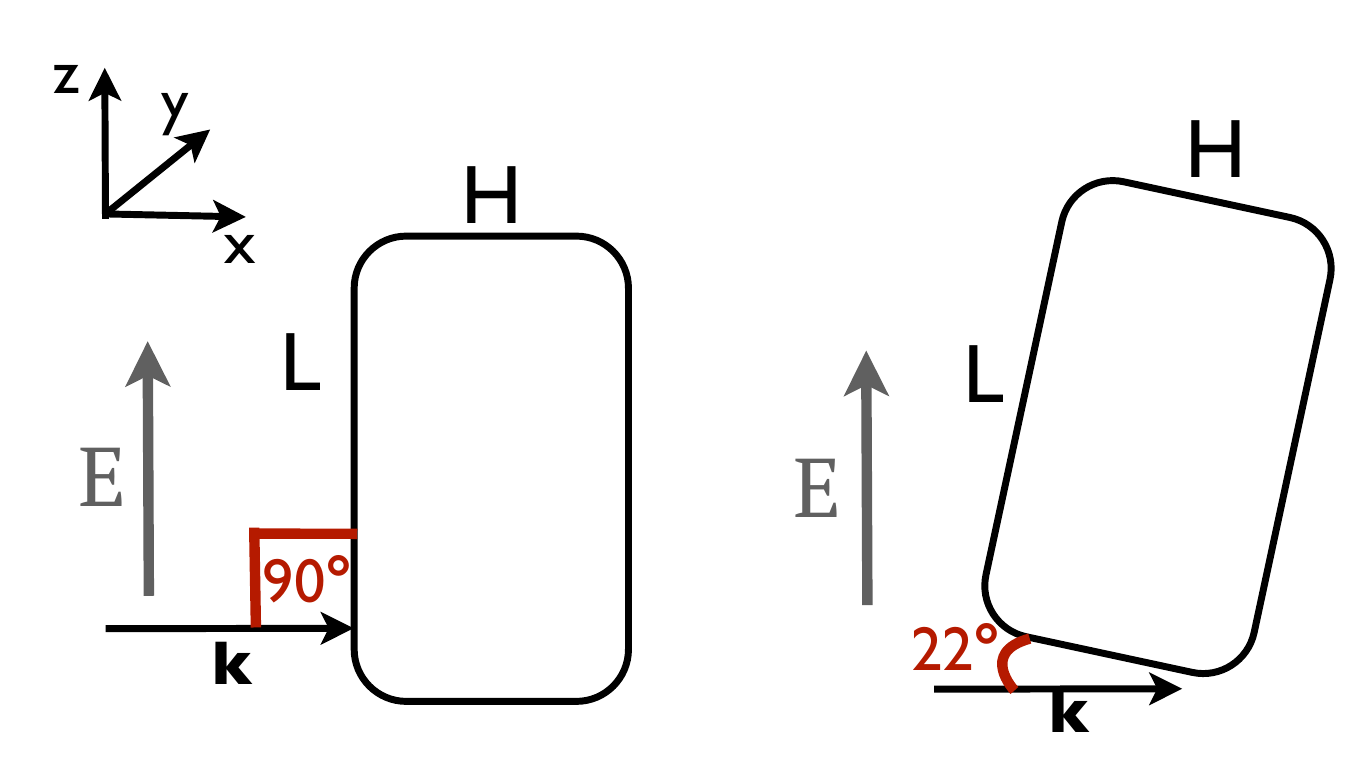} 
    \caption{Diagram showing the angles of incidence in our simulation setups to comply with the configuration in the case from Ellis et al.}
    \label{fig:ellis_ang_inc}
 \end{figure}

\begin{figure}
    \centering
    \includegraphics[width=0.8\textwidth]{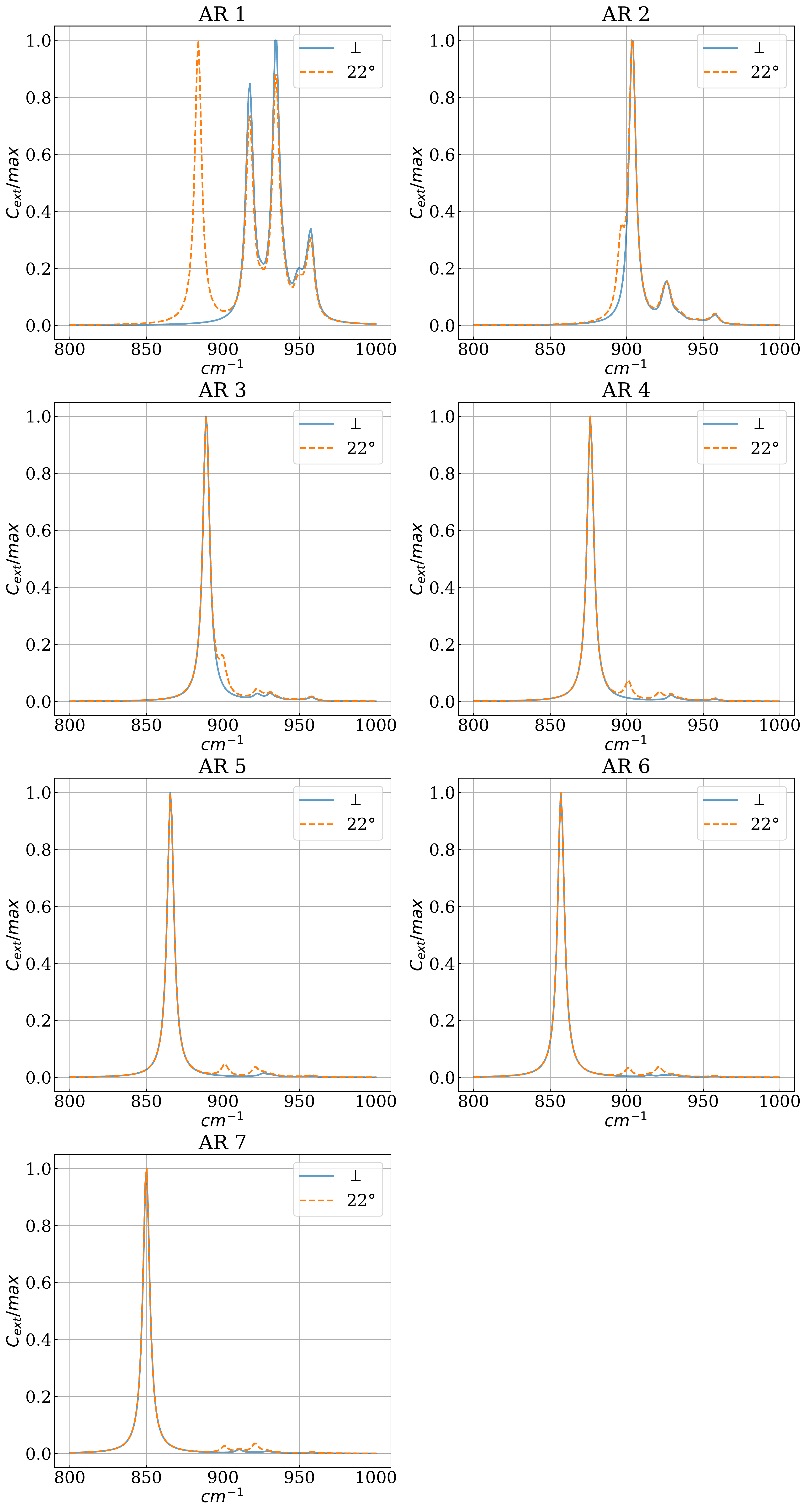} 
    \caption{Extinction cross-section across wavenumbers for SiC pillars of varying aspect ratios,  
             ($H=950$ nm, $W=400$ nm, $L=400$--$2800$ nm, $AR=1$--$7$), with both normal incidence and a 
             22-degree incidence.
            }
    \label{fig:AR_22_vs_norm}
 \end{figure}

\begin{figure}
    \centering
    \includegraphics[width=0.85\textwidth]{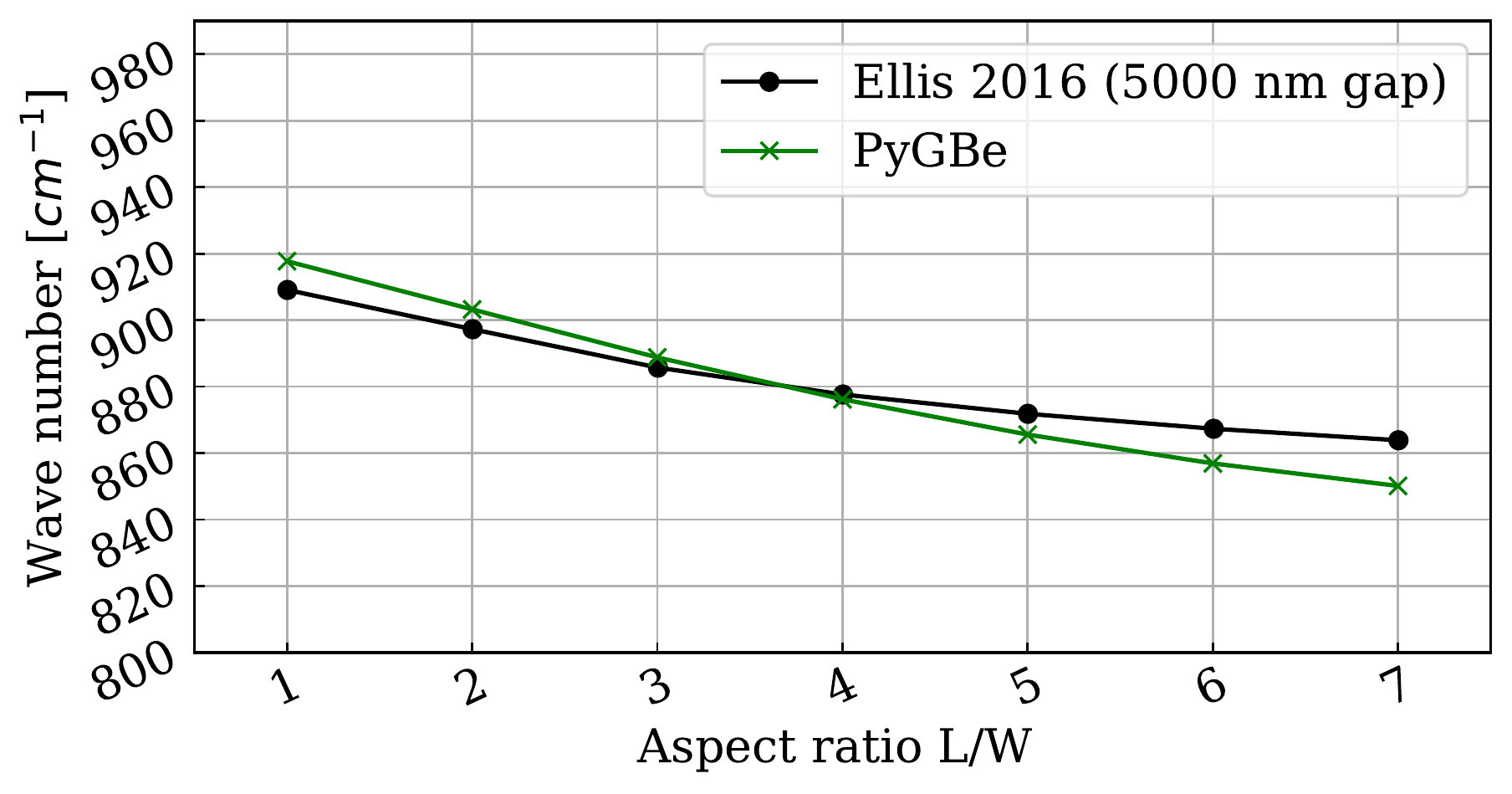} 
    \caption{Replication of figure S4 in the supplementary materials of Ellis et al., 2016. Wave
    number at which the $E^{\parallel}_{100}$ mode happens for different aspect ratios.}
    \label{fig:rep_FS4_ellis}
 \end{figure}

To be able to replicate the result of Figure S4 of the supplementary materials of Ellis et al., we
need to identify the lower-frequency mode for each of
the aspect ratios in our computations. For each aspect ratio ($AR$) 
value from 1 to 7, we computed the extinction cross section $C_{ext}$ across the wave numbers in the range
800--1000 cm$^{-1}$. We identified the lower-frequency mode ($E^{\parallel}_{100}$ for Ellis et al.) that is not a 
longitudinal mode. The longitudinal modes appear only when we have an incidence angle that is off-normal, since 
they are associated with the height of the pillar. Figure \ref{fig:AR_22_vs_norm} shows the results of the extinction 
cross-section of a SiC pillar of fixed height ($H=950$ nm), fixed width ($W=400$ nm) and varied length ($L=400$--$2800$ nm), for normal 
and 22-degrees angle of incidence. The simulations where performed 
for the long-edge orientation, meaning that the electric field is aligned with the length of the pillar 
when having normal incidence as shown on Figure \ref{fig:ellis_ang_inc}. From these results, we selected 
the lowest mode that it is not a longitudinal one and extracted its wavelength (see \ref{tab:ar_peaks}) to replicate 
\ref{fig:rep_FS4_ellis}. Figure \ref{fig:rep_FS4_ellis} shows the results from  Ellis et al. (digitized by hand using the WebPlotDigitizer) and the results obtained
with \pygbe, and Table \ref{tab:err_AR} shows that the percentage error is below 2$\%$ for all cases.
(Note that Ellis et al.\ changed the angle of incidence of the illuminating vector while in \pygbe we 
rotated the geometry instead.) 

\begin{table}
    \begin{center}
      \caption{Wavelength at which peaks happen for different aspect ratios, for runs where the electric
      field is parallel to the length ($L$) of the pillar. We have normal incidence and 22-degree incidence.}
      \label{tab:ar_peaks}
      \begin{tabular}{c c c c c c c c}
        \textbf{AR} \\
        \hline
        \multirow{2}{*}{1} & $\perp$ & \textbf{917.73} & 934.092 & 949.604 & 957.325 \\ 
        & 22$^{\circ}$ & 883.926 & \textbf{917.73} & 935.052 & 949.604 & 957.325 \\ 
        \hline
        \multirow{2}{*}{2} & $\perp$ & \textbf{903.233} & 926.395 & 944.762 & 958.242 \\ 
        & 22$^{\circ}$ & 896.517 & \textbf{903.233} & 926.395 & 944.762 & 958.242 \\ 
        \hline
        \multirow{2}{*}{3} & $\perp$ & \textbf{888.793} & 922.552 & 931.223 & 948.613 & 958.242 \\ 
        & 22$^{\circ}$ & \textbf{888.793} & 899.418 & 922.552 & 931.223 & 958.242 \\ 
        \hline
        \multirow{2}{*}{4} & $\perp$ & \textbf{876.186} & 929.32 & 946.639 & 958.242 \\ 
        & 22$^{\circ}$ & \textbf{876.186} & 901.281 & 921.618 & 929.32 & 945.745 & 958.242 \\ 
        \hline
        \multirow{2}{*}{5} & $\perp$ & \textbf{865.576} & 926.395 & 945.745 & 958.242 \\ 
        & 22$^{\circ}$ & \textbf{865.576} & 901.281 & 921.618 & 958.242 \\ 
        \hline
        \multirow{2}{*}{6} & $\perp$ &  \textbf{856.904} & 914.793 & 923.489 & 929.32 & 946.639 & 958.242\\ 
        & 22$^{\circ}$ & \textbf{856.904} & 901.281 & 920.6 & 958.242\\ 
        \hline
        \multirow{2}{*}{7} & $\perp$ &  \textbf{850.134} & 910.963 & 921.618 & 928.372 & 946.639 & 958.242 \\ 
        & 22$^{\circ}$ & \textbf{850.134} & 901.281 & 910.963 & 920.6 & 958.242\\ 
        \hline
      \end{tabular}
    \end{center}
  \end{table}

 \begin{table}
    \centering
    \caption{Percentage error for different aspect ratios.} 
    \label{tab:err_AR}
    \begin{tabular}{c c}
    \hline
    AR & \% error \\
    \hline
     $1$ & $0.95$ \\
     $2$ & $0.67$ \\
     $3$ & $0.35$ \\
     $4$ & $0.16$ \\
     $5$ & $0.72$ \\
     $6$ & $1.20$ \\
     $7$ & $1.59$ \\
    \hline
    \end{tabular}
\end{table}

\subsubsection{Validation of \pygbe against experimental results in Fig.\ 2a  of Ellis et al., and replication of the corresponding computations}

The results of Figure 2a in Ellis et al.\ were obtained on a pillar of aspect ratio $AR=4$. For this case,
we have the original mesh, provided to us by the authors. We also know that our computation 
for the mode $E^{\parallel}_{100}$ compares well with theirs (percentage error 0.16$\%$), therefore we
attempted to validate our simulations with their experimental results (red curve on their paper), as well 
as to replicate their simulations (green curve on their paper).
Figure 2a of Ellis et al., presents measured and simulated reflectances of SiC pillar
arrays with a 500 nm gap. All their measurements and simulations were performed with 
22$^\circ$ off-normal angle of incidence and incoming polarization parallel to the 
elongated size of the pillar. 

\begin{figure}
    \centering
    \includegraphics[width=0.85\textwidth]{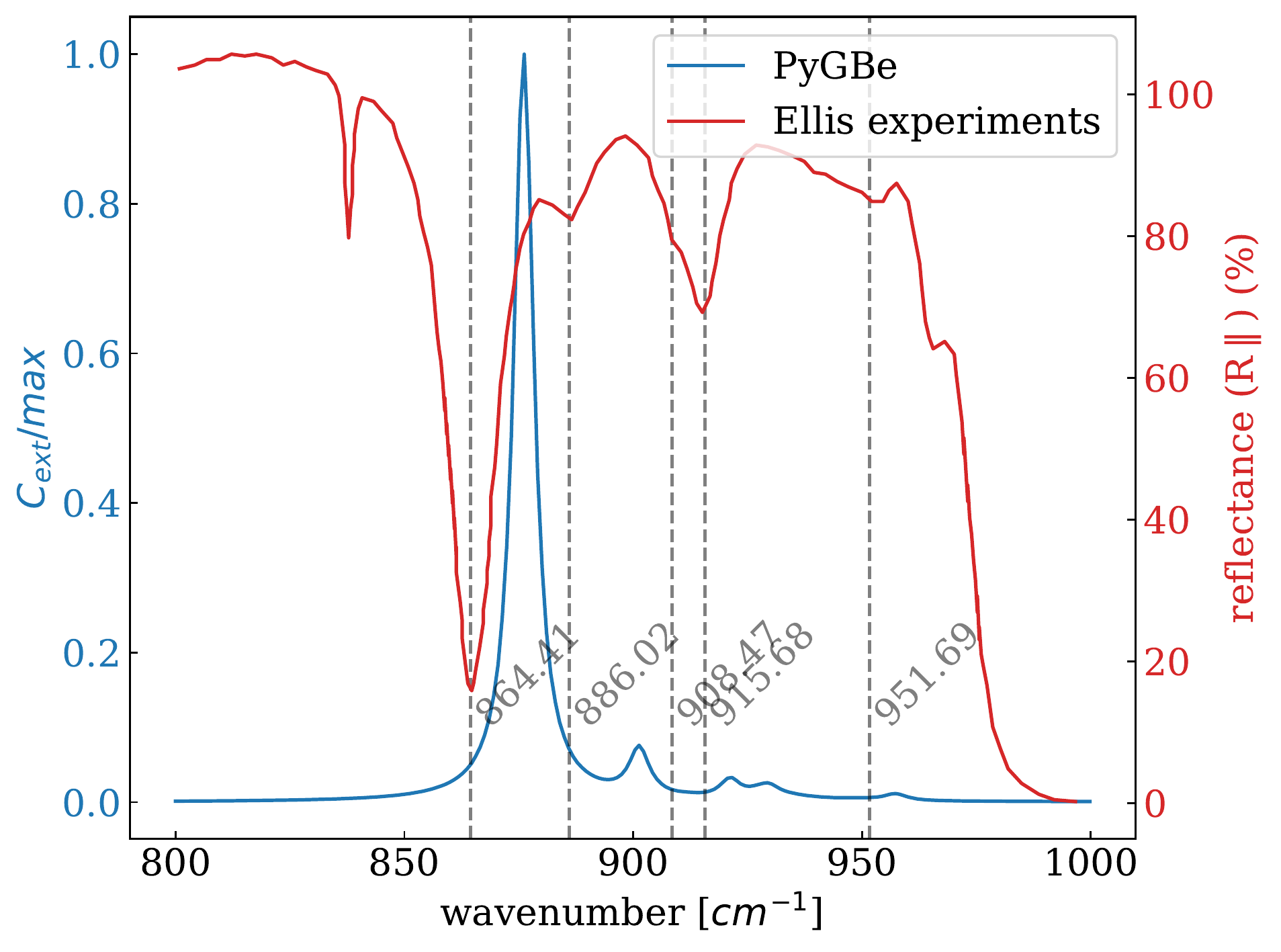} 
    \caption{\pygbe vs.\ the experiments presented in Figure 2a of Ellis et al., 2016 (we obtained the data 
    digitizing by hand from their figure using WebPlotDigitizer).}
    \label{fig:pygbe_vs_exp_2a}
 \end{figure}

Using \pygbe, we computed the extinction-cross section of an isolated SiC pillar ($AR=4$)
with no substrate, submerged in air under a constant electric field in the $z$-direction, 
and rotated the orientation of the pillar to match the angle of incidence used by Ellis et al.\ (see Figure \ref{fig:ellis_ang_inc}).
In Figure \ref{fig:pygbe_vs_exp_2a}, we present a 
comparison of our simulations and the experimental results of Ellis et al. A 
difference in the wavelengths of the peaks is noticeable. This may be attributed to the
fact that in their experiments the separation between pillars is of 500 nm, which implies
there are coupling effects that in our simulations are not considered.  

\paragraph{First-order correction.}

\begin{figure}
    \centering
    \includegraphics[width=0.85\textwidth]{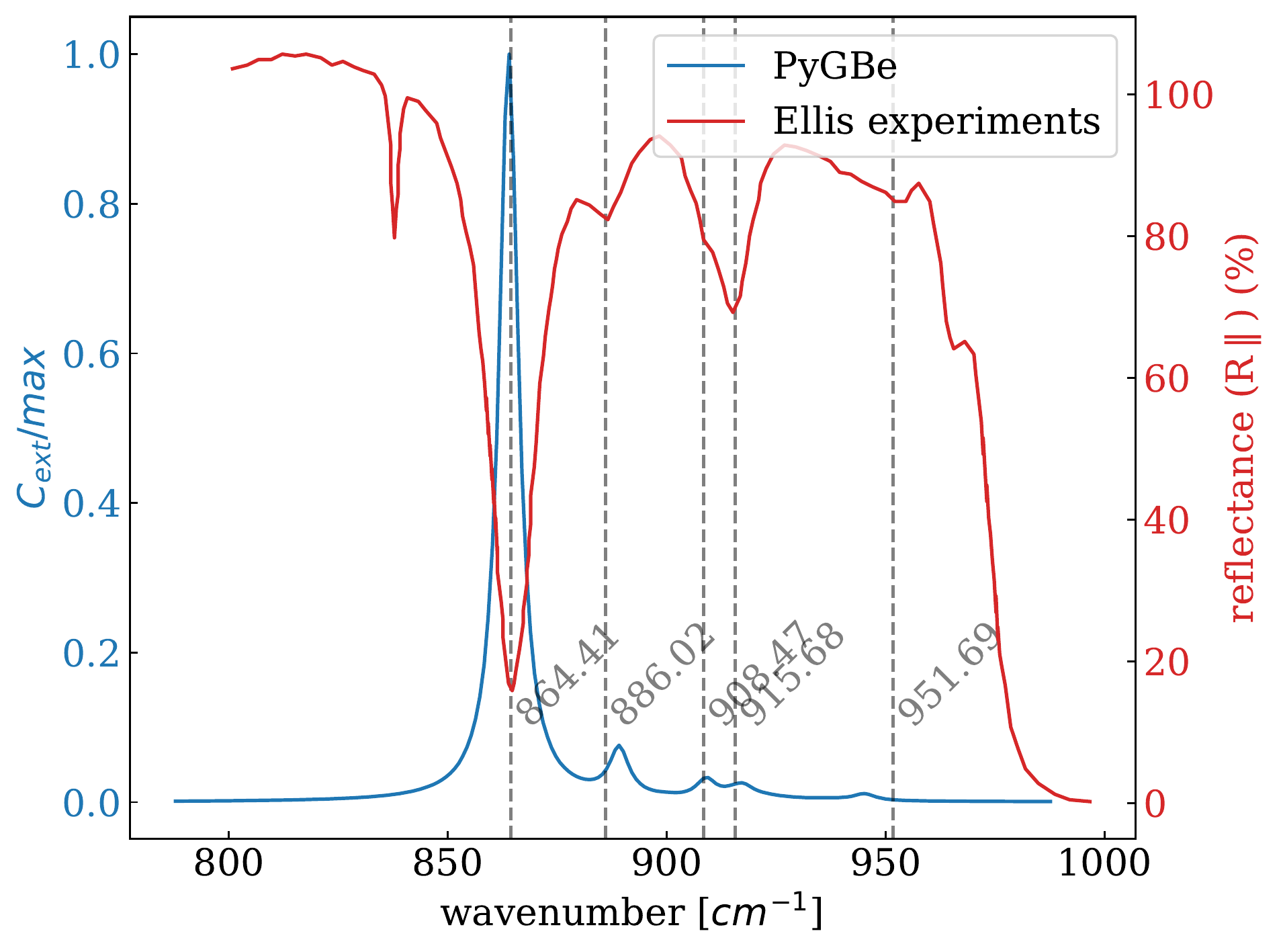} 
    \caption{Validation against experiments in Figure 2a of Ellis, et al., 2016, using the first-order correction, as explained in the text.}
    \label{fig:val_2a}
 \end{figure}

\begin{figure}
    \centering
    \includegraphics[width=0.85\textwidth]{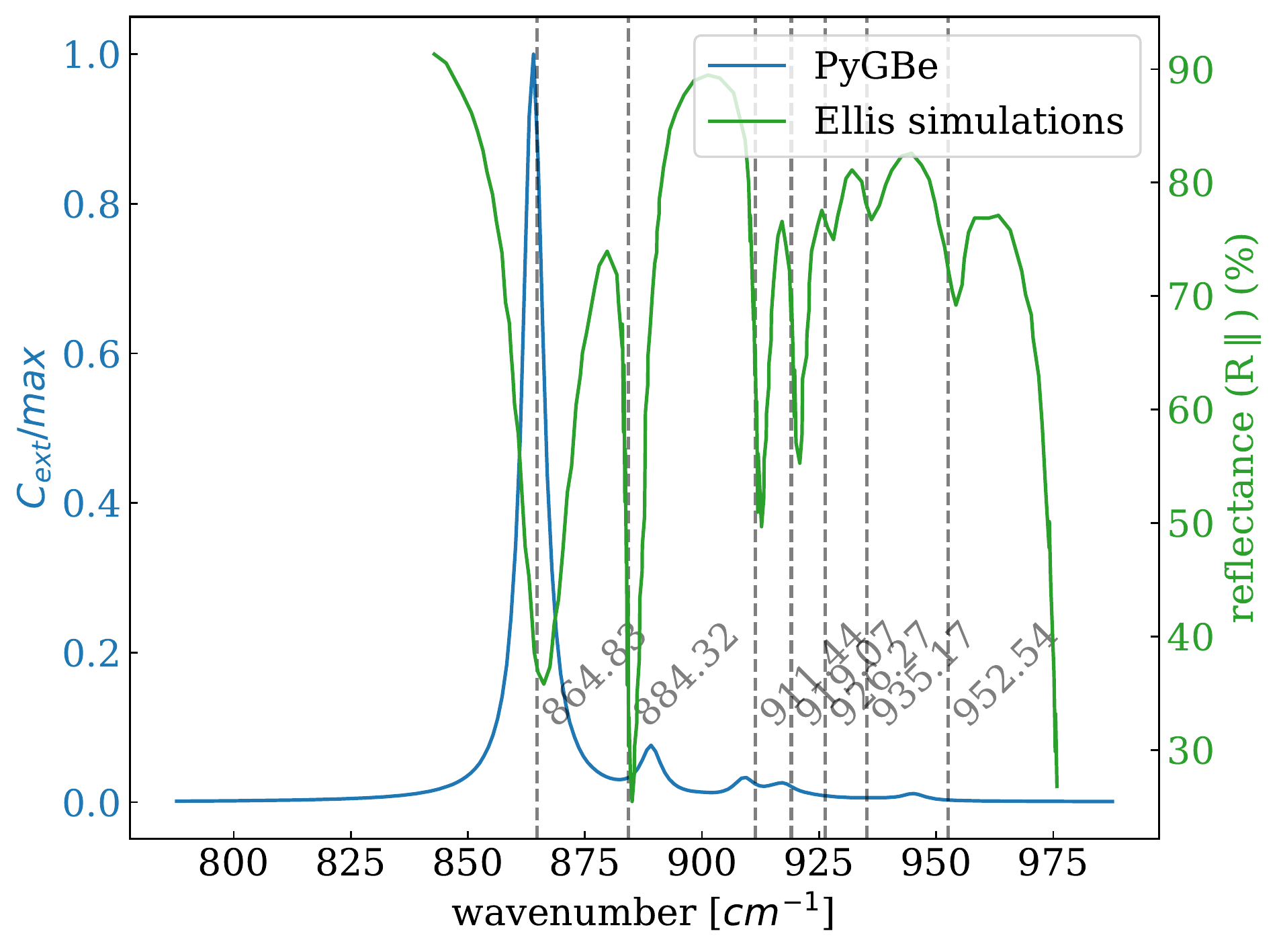} 
    \caption{Replication of the simulations in Figure 2a of Ellis et al., 2016, using the first-order correction, as explained in the text.}
    \label{fig:rep_2a}
 \end{figure}

Since our simulations do not take into account coupling effects in an array of prisms, we cannot strictly match 
the conditions to validate our solver. However, from Figure S4 in the supplementary 
materials of Ellis et al., we know that coupling effects affect the $E^{\parallel}_{100}$ 
mode by a shift of 12.17 cm$^{-1}$. Therefore, as a \textit{first-order correction} we can 
subtract this amount from our simulations to account for coupling effects. In Figure 
\ref{fig:val_2a}, we present the result after applying this correction for coupling effects.
It is worth mentioning that the far-left (837 cm$^{-1}$) and far-right (964 cm$^{-1}$) peaks 
on the results of Ellis et al., are peaks associated with zone-folded LO (longitudinal) phonons of 4H-SiC,
an effect they say is beyond the scope of their analysis \cite{ellis2016}. They concentrate their analysis on the peaks that occur between 864 and 961 cm$^{-1}$.
If we use the same approximation and compare the results with the simulations of Ellis et al.\ on
Figure 2 of their paper (green curve), we obtain Figure \ref{fig:rep_2a}.

 \subsection{Reproducibility and data management} \label{sec:reprod}
 
 Barba (2019) describes the elements of reproducible computational research that we have adopted in our practice \cite{barba2019praxis}. 
 A key element is professional software management and engineering, and the central technology solution is version control. 
 Preserving a complete history of changes is the only way to manage complex software projects that can support reproducibility. 
 Moreover, all our research software is developed in the open, and shared under permissive public licenses, such as BSD-3 or MIT License. 
 These open licenses permit all uses, as well as derivative works, only subject to attribution.
 Another key element of transparent and reproducible computational modeling is \emph{automation}.
Automating every step means \emph{turning protocols into code.}
For example, simulations are launched with parameters fed from a configuration file rather than an interactive prompt, and all figures and plots are produced by writing scripts, rather than pointing-and-clicking on a graphical interface. 
The goal is to create complete ``recipes'' in code, which can be run, versioned, and shared. 

Our signature reproducibility practice is to organize and share in an archival-quality repository (providing a global identifier) all digital artifacts associated with the results in the paper. 
This includes input files, configuration files, post-processing scripts, files specifying the build process and containerization (e.g., Docker files), and even cloud computing machine definitions, if applicable \cite{mesnard-barba2019}.
We call these openly shared file sets \emph{reproducibility packages}, or repro-packs for short, and we have been doing this for many years and improving the process iteratively. 
The basic steps were already contained in the ``Reproducibility PI Manifesto'' of 2012 \cite{barba2012manifesto}, where Barba pledged to always share a manuscript's data, plotting scripts and figures under CC-BY (Creative Commons Attribution license).
Notably, this makes the figures re-usable by readers, without requiring them to ask permissions from the publisher, even if there was a transfer of copyright of the paper (the figures are included in the paper under the conditions of the public license). 
We later extended the practice to include all other digital artifacts associated with the results, beyond secondary data and figures. 
As in our previous papers, readers can reproduce all the figures in this paper using the repro-packs shared in the manuscript GitHub repository, and archived separately in Zenodo. 
They include Jupyter notebooks with all the plotting code, and also the manually digitized data from the figures in the source articles for our replication cases.
Barba and Thiruvathukal, 2017 \cite{BarbaThiruvathukal2017}, explain that it is \emph{not enough} to provide these materials in a GitHub repository (the owner of a repository is always able to delete it), and one should deposit the reproducibility packages in an archival service providing a global identifier and permanent link (like a digital object identifier, DOI).

The following items of archived digital artifacts accompany this paper:

\begin{compactitem}

\item[$\triangleright$] The software repository for \pygbe is at \href{https://github.com/barbagroup/pygbe}{https://github.com/barbagroup/pygbe}.

\item[$\triangleright$] The repository for this paper is at \href{https://github.com/barbagroup/pygbe_validation_paper}{https://github.com/barbagroup/pygbe\_validation\_paper}, which also includes the manuscript source files in LaTeX.

\item[$\triangleright$] The problem datasets for replication of Rockstuhl et al., 2005, are in the manuscript repository, but also archived in Zenodo  at \href{https://doi.org/10.5281/zenodo.3962534}{10.5281/zenodo.3962534}  \cite{ClementiBarba2020-Zen_a}.

\item[$\triangleright$] Execution files for all runs are archived in Zenodo at \href{https://doi.org/10.5281/zenodo.3962576}{10.5281/zenodo.3962576} \cite{ClementiBarba2020-Zen_b}.

\item[$\triangleright$] The problem datasets for validation and replication of results from Ellis et al., 2016 are archived in Zenodo at \href{https://doi.org/10.5281/zenodo.3962584}{10.5281/zenodo.3962584}\cite{ClementiBarba2020-Zen_c}.

\item[$\triangleright$] The file sets for reproducing the figures for the replication of results from Rockstuhl et al., 2005 are archived in Zenodo at \href{https://doi.org/10.5281/zenodo. 3962791}{10.5281/zenodo. 3962791} \cite{ClementiBarba2020-Zen_d}.

\item[$\triangleright$] The file sets for reproducing the figures for the validation and replication of results from Ellis et al., 2017 are arhived in Zenodo at \href{https://doi.org/3962797/zenodo.3962791}{10.5281/zenodo.3962797} \cite{ClementiBarba2020-Zen_e}.

\end{compactitem}

\section{Discussion}\label{sec:discussion}

We have presented several results that replicate previous published findings in the general area of nanostructure responses to electromagnetic waves. 
Our field of interest is computational nanoplasmonics for applications in biosensors, and in a previous publication we developed the mathematical formulation and reported both solution verification activities, and an application demo with our software \pygbe, extended to treat complex dielectrics and imposed electric fields \cite{ClementiETal2019}.
The search for a physical context and published results that would allow us to undertake validation studies with \pygbe is what led us to this work. 
Even if we finally do have validation and replication cases, neatly presented here, the path to obtain these results was nonlinear, iterative, and arduous.

In the first case, we sought to replicate a result from Rockstuhl et al., 2005 \cite{rockstuhl2005}, where they computed the scattering cross-section as a function of wavelength for a silicon carbide rectangular nanostructure. 
They present their results as a plot (Figure 14 in their paper), and report the numeric value of the resonance wavelengths in the text. 
Lacking access to the secondary data behind the plots---computed from two-dimensional simulations with a boundary element solver---we were forced to manually digitize the values from the figure.
Our results are presented in Figure \ref{fig:rep_14}, together with the curve we obtain from digitizing the source image. 
We were successful at replicating the strong peaks reported  by Rockstuhl et al.\ at wavelengths
10.42 $\mu$m and 10.7 $\mu$ m,  when the electric field $E$ is parallel to the short side of the rectangle, and 10.42 $\mu$m and 10.82 $\mu$ m
when $E$ is parallel to the long side. Our results contain extra (small) peaks that are not present in the work of Rockstul et al.
The first one, located between the main two peaks, we attribute to the the effect of
sharp edges (see Figure \ref{fig:tri_reg_round_14}), as when we introduce a level of roundness, it diminishes. The second extra peak is the far right one, and we believe this
peak is a consequence of the 3D nature of our geometry; as observed in Figure \ref{fig:ext_y_14} this peak intensity
decreases as the third dimension of the geometry lengthens.
The quantity of interest in these findings is the wavelength of the resonance peaks, and our results do indeed match the findings.

The second replication case is from a paper by Ellis and co-workers \cite{ellis2016} studying the effect of aspect ratio on the excitation of high-order modes in localized surface phonon-polariton nanostructures (Figure S4 of their supplementary material and Figure 2 of their paper).
Figure \ref{fig:rep_FS4_ellis} shows the results for this replication, where the relative errors between our computations and theirs is always smaller than 2$\%$. The smallest error is for the case with $AR=4$ and we proceeded to study the results on Figure 2 of their work, corresponding to this same aspect ratio.
Their results in Figure 2a include both experiments and simulations with the commercial software COMSOL, so we sought to both validate \pygbe using their experimental results, and replicate their computational findings. 
Again, we lack access to the data behind the plots, and we had to digitize the curves by hand. 
The quantity plotted in the original figures is reflectance as a function of wavenumber, whereas we compute the extinction---on the figures, they show inverted peaks, where we show positive peaks.
We can compare the results, nevertheless, because the quantity of interest is the wavenumber position of the peaks.
Figure \ref{fig:pygbe_vs_exp_2a} shows the results of our simulations using \pygbe on an isolated pillar,
compared with the experimental results of Ellis et al.\ on an array of pillars.
The results with \pygbe  do not account for the effect of coupling among the pillars, which explains the noticeable discrepancy on the wave numbers at which the peaks occur. 
We proposed a correction, based on the results reported on Figure S4 of Ellis et al.\ for the $AR=4$ case, where the shift on the wave number due to coupling is 12.17 cm$^{-1}$ (difference between black and red curves for $AR=4$ on figure S4 of the supplementary materials). 
We subtracted this value to that obtained with our simulations and we show the comparison of our corrected results against those of Ellis et al.: the experimental data on Figure \ref{fig:val_2a}, and the
computational data on Figure \ref{fig:rep_2a}. 
When comparing with their experiments, after the correction was applied, we observe a good match of the wavenumber for the lower (and stronger) mode, as well as a good match for the 
third and fourth peaks. The wave number of the second peak, related to a longitudinal excitation (mode $L_{000}$ in Ellis et al.), presents a discrepancy that we believe is related to the fact that our 
pillar does not have a substrate underneath. The remaining (fifth) peak, also presents a discrepancy, but in this case we can not identify the reason.
We did not analyze the peaks out of the range 864--961 cm$^{-1}$, since Ellis et al. \ describe these peaks to be associated with zone-folded LO (longitudinal) phonons of 4H-SiC,
and outside the scope of their study.
After considering all these details, we can say that we have validated our solver \pygbe against the experimental results of Figure 2a, as well as replicated their computational results.


Throughout all these activities aiming to replicate previous results and validate our computational model, we faced multiple challenges, starting with the complexity of the system under study. In both our source papers, we have systems that we were unable to fully 
model using \pygbe. For the case of Rockstuhl et al., even though they used a boundary element method, it was a 2D model instead of 3D like the one implemented in \pygbe. The computational work presented in Ellis et al.\ used a volumetric 
formulation (finite element method), whereas \pygbe implements a boundary integral one. Both works computed (or measured) quantities different from the extinction cross-section (computed in \pygbe), but given he relationship between their quantities (scattering cross-section and reflection)  
and ours (extinction cross section), via the wavelength (wavenumber) at which resonance happens, this was not a problem. To produce the end results for comparing against these works, however, we went through an 
exhaustive process of modeling, making assumptions, and even corrections. We were lacking any information regarding the solvers, discretization, meshes, etc., in the original papers. In the case of Ellis et al., we benefited from multiple communications with the authors, 
who made available the geometry and mesh of the pillar, as well as the dielectric data used as input for the simulations. Even though we appreciate this helpful interaction, replication studies should ideally not depend on communications with the original authors. 
In both cases, the secondary data presented in their results were not publicly available and we relied on a manually digitized version, obtained from the plots. This work is tedious and introduces uncertainties that would be avoided by releasing 
the secondary data.

Validation has been a hard goal to achieve for our solver since the field does not have benchmarks that are meant to be used for validation. Multiple times,
we encountered experimental results that we aspired to replicate, but due to the insufficient reporting of the experimental conditions, the setup and the geometries involved, 
we have been unable to validate until this moment. Even though we were not modeling the same experimental setup, the communications with the authors of Ellis et al., made 
the validation possible.

After all the challenges faced to achieve code validation and replication of results, we can say that the process was far from  linear, 
and the complexity involved increases if the work to be replicated or used for validation was not conducted and published using reproducible practices, and made open at the time 
of publication. We conclude that reproducible practices are needed not only for the work to be reproduced, but also 
replicated and even used for validation studies, if applicable.

\section*{Acknowledgments}

We thank Dr. Chase T. Ellis  and Dr. Joseph G. Tischler for the fruitful discussions regarding their results \cite{ellis2016}, 
which led to the validation of \pygbe, and for providing a mesh file and dielectric data that made the modeling closer to their original setup.
We also thank Dr. Christopher D. Cooper for many valuable discussions that helped us overcome some of the challenges faced throughout this work. 

This research is supported by the National Science Foundation under Award \#1747669 of the Computer and Information Science and Engineering (CISE) Directorate.

\bibliographystyle{unsrt}
\bibliography{pygbe_rep_val} 

\end{document}